\documentclass[aps, nobibnotes, secnumarabic, showpacs, amssymb, amsmath,
prb, twocolumn ]{revtex4}

\usepackage{epsfig}
\usepackage{endnotes}
\let\footnote=\endnote

\begin{document}

\def\p{\phi}\def\P{\Phi}\def\a{\alpha}\def\e{\epsilon}\def\be{\begin{equation}}
\def\ee{\end{equation}}\def\l{\label}\def\0{\setcounter{equation}{0}}
\def\b{\beta}\def\S{\Sigma}\def\C{\cite}
\def\r{\ref}\def\ba{\begin{eqnarray}}\def\ea{\end{eqnarray}}
\def\n{\nonumber}\def\R{\rho}\def\X{\Xi}\def\x{\xi}\def\la{\lambda}
\def\d{\delta}\def\s{\sigma}\def\f{\frac}\def\D{\Delta}\def\pa{\partial}
\def\Th{\Theta}\def\o{\omega}\def\O{\Omega}\def\th{\theta}\def\ga{\gamma}
\def\Ga{\Gamma}\def\t{\times}\def\h{\hat}\def\rar{\rightarrow}
\def\vp{\varphi}\def\inf{\infty}\def\le{\left}\def\ri{\right}
\def\foot{\footnote}\def\vep{\varepsilon}\def\N{\bar{n}(s)}
\def\k{\kappa}\def\sq{\sqrt{s}}\def\bx{{\mathbf x}}\def\La{\Lambda}
\def\bb{{\bf b}}\def\bq{{\bf q}}\def\cp{{\cal P}}\def\tg{\tilde{g}}
\def\cf{{\cal F}}\def\bN{{\bf N}}\def\Re{{\rm Re}}\def\Im{{\rm Im}}
\def\bk{\hat{\mathbf{k}}}\def\cl{{\cal L}}\def\cs{{\cal S}}\def\cn{{\cal N}}
\def\cg{{\cal G}}\def\q{\eta}\def\ct{{\cal T}}\def\bbs{\mathbb{S}}
\def\bU{{\mathbf U}}\def\bE{\hat{\mathbf e}}\def\bc{{\mathbf C}}
\def\vs{\varsigma}\def\cg{{\cal G}}\def\ch{{\cal H}}\def\df{\d/\d }
\def\mz{\mathbb{Z}}\def\ms{\mathbb{S}}\def\kb{\hat{\mathbb
K}}\def\cd{\mathcal D}\def\mj{\mathbf{J}}\def\Tr{{\rm Tr}}
\def\bu{{\mathbf u}}\def\by{{\mathrm y}}\def\bp{{\mathbf p}}
\def\k{\kappa} \def\cz{{\mathcal Z}}\def\ma{\mathbf{A}}
\def\me{\mathbf{E}}\def\ra{\mathrm{A}}
\def\ru{\mathrm{u}}\def\rP{\mathrm{P}}\def\rp{\mathrm{p}}\def\z{\zeta}
\def\my{\mathbf Y}\def\ve{\varepsilon}\def\bw{\mathbf
W}\def\hp{\hat{\p}}\def\hh{\hat{h}}\def\hx{\hat{\x}}\def\hk
{\hat{\kappa}}\def\hj{\hat{j}}\def\eb{\mathbf{e}}\def\bj{\mathbf{j}}
\def\tO{\tilde\o}\def\cS{{\cal S}}\def\os{\overline{\s}} \def\K{\hat{K}}
\def\mH{\mathcal{H}} \def\he{\hat{e}}
\def\to{\rightarrow}\def\t{\times}\def\u{\hat e}\def\ep{\epsilon}

\begin{flushright}{Dedicated to Alexei Sissakian}\end{flushright}

\title{On the Generalized Correspondence Principle}
\author{J.~Manjavidze}\email{joseph@jinr.ru}
\affiliation{ Andronikashvili Institute of Physics, Tbilisi State
University, Georgia,} \affiliation{Joint Institute for Nuclear
Research, Russia}

\begin{abstract}

It is noted that quantum theory includes the correspondence principle
which is independent of the value of quantum corrections. Some
consequences of this generalized correspondence principle are
considered. For example, it is discussed how the symmetry constraints
can be taken into account.

\end{abstract}


\received{}

\maketitle

{\bf 1.} The aim of the article is to allow readership to take a look
at the generalized correspondence principle (GCP): {\it the quantum
process probability is defined by equilibrium among ordinary
mechanical forces and quantum force $\hbar j$ of the arbitrary
strength if a classical analog exists} \C{1} \footnote{ Another
formulation of GCP also exists: {\it for an external observer the
quantum system looks like classical which is excited by quantum force
if a classical analog exists.} This interpretation was offered by
A.Sissakian.}. Notice that GCP is close to the principle of
d'Alembert: {\it the dynamics of the system is defined by equilibrium
of all acting in it forces, if the motion is time reversible} \C{2}.
The d'Alembert principle was formulated in 1743, long before the
quantum mechanics but by all appearance it has the general sense.

At the same time GCP differs from the well known Bohr correspondence
principle offered in 1923 which runs that {\it the quantum mechanical
systems behavior can be described by classical mechanics laws in the
limit of large quantum numbers} \C{3}. The role of the Bohr
correspondence principle can not be overestimated. It introduces the
Newton laws into the quantum mechanics: if there is a classical
analog in the system then the {\it in}-state obtains the exponential
factor $e^{iS/\hbar}$ during its transition into the {\it out}-state
according to the general conclusion of P. Dirac \C{4}. It is
important here that $S$ have the structure of the classical action.

Let us consider a particle collision process. The point is that the
quantum nature of collision leads to the perturbation of the object
in question as well as of the incident particle. One of the possible
scenario of the process may be marked by the given field $\vp(x)$,
where $x$ is the 4-coordinate in Minkowski metric, $x^2=t^2-\bx^2$.
This gives the factor $e^{iS(\vp)/ \hbar}$ to each configuration of
$\vp$. Therefore quantum nature of the problem means the necessity to
take into account all configurations, i.e. to find the sum (integral)
over $\vp$ \C{5},
$$A=\Sigma_{\{\vp\}} e^{iS(\vp)/\hbar}.$$ The definition of set
$\{\vp\}$ is a fundamental problem of quantum theories and this
problem is solved by GCP. The Bohr correspondence principle partly
solves it: the variation of $S(\vp)$ over $\vp$ must be equal to
zero, \be \f{\d S(\vp)}{\d\vp(x)}=0\l{1}\ee in the limit $\hbar=0$.
That equation simply means that it is desirable to have the minimal
oscillations during the transition $(in\to out)$ state. It is held in
quantum case with exponential accuracy and Eq. (\r{1}) becomes exact
only in the classical limit $\hbar=0$. Eq. (\r{1}) means the
equilibrium among ordinary kinetic and potential forces in terms of
d'Alembert.

{\bf 2.} The key word for GCP is the probability, $R$, which is equal
to modulo square of amplitude $A$, $R=|A|^2=A\cdot A^*$. Notice first
of all that only the absorbed wave can be detected by an experiment
since the total probability is a conserved quantity, i.e. following
the optical theorem, $A\cdot A^*=2\Im A$, only the imaginary
(absorption) part of amplitude, $\Im A$, describes the result of the
measurement. Notice also that the product $A\cdot A^*$ describes the
time reversible (a closed path) motion since the end points of the
motion $(in \to out)$ which is described by $A$ and $(in\leftarrow
out)$ described by $A^*$ coincide. As a result the phase of $A$ is
canceled in the modulo square $R=|A|^2$. Therefore, the notion of
quantum probability contains extra requirements and GCP is their
consequence \footnote{We will assume that the interaction with
external fields switched on adiabatically}.

For example, let us consider a motion of a particle with energy $E$
in a potential hole. Corresponding amplitude looks as follows:
$$ A(\bx_1, \bx_2;E)=\sum_n \f{\psi_n(\bx_1)\psi^*_n(\bx_2)}
{E-E_n+i\ve},$$ where $E_n$ is the bound state energy. One can note
that $A(\bx_1, \bx_2;E)$ is defined on the whole axis of $E$. Let us
consider now the probability $\R(E)$ to find the particle with energy
$E$. Taking into account the ortho-normalizability of the wave
function $\psi_n$ one can find that $$\R(E)=\int d\bx_1d\bx_2
|A(\bx_1,\bx_x;E)|^2=$$\be=\sum_n \f{1}{|E-E_n+i\ve|^2}=\f{\pi}{\ve}
\sum_n\d(E-E_n).\l{*}\ee It is noticeable also that $\R(E)\neq0$ only
for physically acceptable values of $E=E_n$. One may conclude that
the rightly defined observable contains the above mentioned
requirements inherently.

One can show \C{1} that (i) if a classical analog in the system
exists \footnote{In other words, if the correspondence principle of
Bohr is applicable in the limit $\hbar=0$.}, (ii) if $S$-matrix is
unitary and (iii) because of the closed-path boundary conditions we
have: \be R=\lim_{j=e=0}e^ {-i\kb(j,e)} \int DM
e^{iU(\vp,e)/\hbar},\l{3}\ee where $\kb$ in the first exponent is the
perturbations generating operator, $$ 2\kb=\int dx \f{\d}{\d
j(x)}\f{\d}{\d e(x)},$$ the interactions are induced by the
functional $$ U(\vp,e)=S(\vp+e) - S(\vp-e)- 2\int dx \f{\d
S(\vp)}{\d\vp(x)}e(x),$$ where $e=e(x)$ is the auxiliary field which
describes the quantum deformations of $\vp$, and the Dirac measure
$$ DM=\prod_x d\vp(x)\d\le(\f{\d S(\vp)}{\d\vp(x)}+ \hbar j(x)\ri).$$
One can trace the analogy of $\d$-function in $DM$ with $\d$-function
which has appeared in Eq. (\r{*}). Therefore because of the Dirac
$\d$-function $DM$ is defined on the {\it complete set} of the
physics field configurations. The formula (\r{3}) describes the
vacuum-into-vacuum transition probability. An analogous formula
exists in the case of the particle production \C{7} if the
interaction with external states are switched on adiabatically, i.e.
if the external states has not an influence on the interacting fields
spectrum.

But it must be underlined that it is impossible to use (\r{3}) for
the quantum tunneling effects, when the action $S$ becomes complex
quantity\footnote{The tunneling effects can be described by an
analytical continuation of the classical trajectory under the
potential barrier \C{6}.}. The integral (\r{3}) can be analytically
continued on the complex time contour $C=C_++C_-,~C_\pm:~t\to t\pm
i\ve,~\ve\to+0$, to avoid the singularities if they occur.

It is noticeable that $DM$ is defined by the strict equation (GCP):
\be \f{\d S(\vp)}{\d\vp(x)}=-\hbar j(x). \l{2}\ee It is correct for
arbitrary value of $\hbar$, not only for semiclassical approximation
$\hbar\to0$. Meanwhile the quantum corrections are generated by the
given operator $\kb$.

The reason of Eq. (\r{2}) appearance is as follows. The main point of
the GCP formalism is cancelation of $\Re A$ if $A\cdot A^*$ is
calculated. It becomes possible because of the closed-path
prescription which ensures the equilibrium of forces, i.e. the
dynamics must be defined by Eq. (\r{1}) at $\hbar=0$ (Bohr). The
"additional" quantum force, $\hbar j(x)$, is introduced at
$\hbar\neq0$ to provide equilibrium (d'Alembert). As a result we come
to the $exact$ Eq. (\r{2}).

{\bf 3.} Let us consider now the most evident consequences of GCP.
First of all $\d$-likeness of the measure, $DM$, orders to use a
complete set of strict solutions of Eq. (\r{2}) in vicinity of $j=0$.
Notice there is a missing interference among various trajectories.

Next, only one term in the sum over solutions must be taken into
account. Indeed, let us consider the simplest example of a particle
moving into a potential hole which has a minimum at the origin,
$\vp=0$. Then Eq. (\r{2}) will have two solutions at the classical
limit when the source of quantum perturbations $j=0$: the particle
all time rests on bottom of the hole, $\vp_1=0$, and the particle
moves with energy $\e$, $\vp_2=u(t+t_0, \e)$, where $t_0$ is a
starting time. Turning on $j(t)$ one can find two quantum
trajectories.

One must integrate over $t_0$ and $\e\geq0$ since one must take into
account all trajectories. In this case the contribution of $u(t+t_0,
\e)$ is proportional to the infinite integral over $t_0$, i.e. it is
proportional to the volume of the group of time translations, $\O$.
At the same time the contribution of $\vp_1$ is finite. Therefore, as
it follows from the definition of Dirac $\d$-function, $R=R(\vp_1)
+\O R(\vp_2)$ and the contribution of $\vp_1$, $R(\vp_1)$, is
improbable since it is realized on zero measure: its contribution is
$\sim 1/\O$ in comparison with the contribution of $\vp_2=u$. In
other words the contribution of $\vp_1$ occupies a point-like volume
in the phase space and as the consequence it must be omitted as it is
improbable.

One may conclude that in the situation of general position \C{8},
when there is no external influence, the trajectories which maximally
break the symmetry of the action are most probable. The path-integral
formalism based on this selection rule\footnote{This selection rule
has not been used in the ordinary formulations of quantum theories
but it is widely known in classical mechanics, see for example
formulation of Kolmogorov-Arnold-Mozer theorem in the textbook
\C{8}.} becomes self-consistent, it does not require external
assumptions and leads to {\it the theories with symmetry}. One can
note that the gauge invariant quantum chromodynamics (QCD), just as
quantum electrodynamics (QED), can not be considered as the theory
with symmetry since the space-time symmetry stays untouched in
it\footnote{The gauge degrees of freedom play no role here since an
arbitrary solution of Yang-Mills equation breaks it.}.

The measure of contribution in the general case is defined by a
number of parameters $(\q,\x)$, such as $(\e,t_0)$ in the considered
example: $\q=\e$ and $\x=t+t_0$ in the semiclassical approximation.
They form the space $W$, $(\q,\x)\in W$, and the volume of $W$, $\O$,
defines the measure of a given trajectory. Therefore, our selection
rule means that the trajectories of the largest dimension $W$ are
most probable.

{\bf 4.} Another important property of GCP is a possibility to
perform the transformation of integrants, i.e. to use the most
powerful method of classical mechanics. This becomes possible due to
the $\d$-likeness of the measure: a transformed measure will be again
$\d$-like which ensures conservation of the total probability. A
naive transformation of variables in the amplitude is cumbersome and
usually produces bad results \C{11} due to stochastic nature of
quantum fluctuations \C{12}.

The transformations can be used to formulate the quantization of
non-linear wave $u$ in terms of the generalized coordinates\footnote{
It must be underlined here that the procedure of the accounting of
the symmetry constraints can not be done perturbatively. One can
understand the quantization as a formal language which contains the
words ({\it independent} degrees of freedom) and the rules
(mathematics) which explain how a sentence (result) must be
deduced.}.

The first question is: how can "the generalized coordinates" be
defined in quantum theories\footnote{One can provide following
elegant geometrical interpretation of the "generalized coordinates"
approach. Following the general formulation \C{14, 14a} the set of
parameters $\ga$ forms the factor space $W$ and $u(x;\ga)$ belongs to
it completely. The mapping of multi-particle dynamics into $W$ forms
the finite-dimensional hypersurface in it. The hypersurface
compactifies into the Arnold's hypertorus if the classical system is
completely integrable \C{8}. Then one half of parameters $\ga$ is the
radii, $\q$, of the hypertorus and the other one is the angles, $\x$.
Description of the quantum system in terms of the collective-like
variables $\ga$ means the description of random deformations of this
hypersurface, i.e. of the surface of Arnold's hypertorus in the case
of the integrable system. }? Let us return to the example of a
particle in the potential hole. Instead of the $\vp$ one can use the
{\it energy-time} variables having the non-linear wave $u(t+t_0,\e)$.
The transformation into the space of parameters of classical
trajectory, $W$, is preferable since the classical trajectory $u$ is
defined by the coordinates of $W$ space completely, see also \C{15}.

It has been shown that if a system has the sufficient set of
conserved parameters, i.e. is completely integrable, then new
variables, $(\q,\x)$, form a simplectic manifold, $T^*W$, and all of
them are quantum variables, $q$-numbers. The example of such a system
is sin-Gordon model, where $\q$ is the momentum of soliton and $\x$
is the corresponding coordinate \C{16}. If the completely integrable
system has a hidden symmetry than only a portion of new variables
form the simplectic manifold, $T^*W$, and the others, $\la$, are
ordinary numbers of the zero-modes space, $C$. The example can be
Coulomb problem \C{17}. In it a pair $(angular~momentum,angle)$
belongs to $T^*W$ and the conserved hidden parameter, the length of
$Runge-Lenz~vector$, belongs to $C$. We have the same in the case of
Yang-Mills fields with symmetry \C{18}. Therefore, one may expect
that $W=T^*W+C$.

The transformation of l.h.s. of Eq. (\r{2}) leads to transformation
of r.h.s., $j$, or more exactly to the splitting of $j$ on the
projection on the axes of $T^*W$ space, $j\to (j_\q,j_\x)$. The
variable $e$ must be split also, $e\to (e_\q,e_\x)$, to conserve the
form of perturbations generating operator $\kb$. Therefore, we have
come to the theory like the Hamiltonian quantum mechanics, where
$\q=\q(t)$ may be interpreted as a generalized momentum and
$\x=\x(t)$ as a conjugate to $\q$ generalized coordinate of a
particle and $u(\bx;\q,\x,\la)$ defines the external potential. It
must be noted that $T^*W$ is a uniform and isotropic space in the
semi-classical approximation since $\q$ are time independent
quantities in that case. The quantum theory in $T^*W$ describes
fluctuations disturbing this property.

It has been demonstrated that the programm of reduction of the
quantum field-theoretical problem with symmetry of the arbitrary
dimension to the quantum-mechanical one, \be u:~\vp(\bx,t)\to
(\q(t),\x(t),\la),~ (\q,\x)\in T^*W,~\la\in C,\l{4}\ee can be
realized. It is important that the divergences accompanying (\r{4})
are canceled in the integral (\r{3}) since the total probability is
conserved. As a result we have come to the field theory with symmetry
without any divergences, like quantum mechanics, if $S(u)$ is finite.
Indeed, as the result of mapping $R$ have the same structure (\r{3})
with \be DM=d\la\prod_t d\q d\x \d\le(\dot\q+\f{\pa
h}{\pa\x}+j_\q\ri) \d\le(\dot\x-\f{\pa h}{\pa\q}-j_\x\ri),\l{5}\ee
where $h(\q,\x,\la)=H(u,\dot u)$ and $H(u,\dot u)$ has the structure
of incident Hamiltonian. Notice that we use terms of the Lagrange
formalism up to transformation (\r{4}). Next, the interactions are
described by $U=U(u,e),$ where
$$e=e_\x\f{\pa u}{\pa\q}-e_\q\f{\pa u}{\pa\x}$$ and the
perturbations are generated by \be 2\kb=\int dt\le\{\f{\d}{\d
j_\q}\f{\d}{\d e_\q}+\f{\d}{\d j_\x}\f{\d}{\d e_\x}\ri\}.\l{6}\ee The
singularity of (\r{4}) means that this mapping is irreversible, while
one must stay in $W$ space forever. On the other hand the mapping is
necessary because of the quantization rule: one must work with
independent degrees of freedom. Therefore, the sense of QCD
small-distance effects may be vague in the light of GCP, i.e. if QCD
is realized on the zero measure, and it is possible that they need a
new interpretation\footnote{Physics knows such an example when a
"good" theory was rejected.  Weber's electrodynamics for instance
perfectly describes all experimental facts, except the radiation of
light.}.

{\bf 5.} Moreover, there is no radiation in the field theories with
symmetry without matter fields. Indeed, the operator (\r{6}) on the
measure (\r{5}) generates the {\it complete} set of contributions
which does not contain the particle states\footnote{The free particle
must have a definite energy and momentum in the traditional
understanding of it, see \C{20}.}. In other words the field theory
with symmetry does not contain even the notion of the "particle".
This conclusion is rightful for an arbitrary field theory with
symmetry.

The prove of this qualitative conclusion is simple \C{physrev-1}.
Namely, one can compute the $m$ into $n$ massless particle transition
cross section, $\R_{mn}$, times the flux factor \C{carr} to show that
\be \R_{mn}=0, ~\forall\{m,n\}>0, \l{7}\ee in the field theory with
symmetry. The prove of Eq. (\r{7}) is simple. Thus, following to the
Lehman-Symanzick-Zimmerman reduction formula \be \Ga^*(q)=\int dx
e^{iqx}\pa^2 u(\bx;\x(t),\q(t),\la) \l{8}\ee is the particles
production vertex. Then $\Ga^*(q)=0$ if the field $u=0$ on the
infinitely far hypersurface $\s_\infty$ since $\Ga^*(q)$ stocks up on
$\s_\infty$ as it follows from (\r{8}) and the real particles
production, with $q^2=0$, is considered. GCP quantum corrections can
not change the asymptotics of $u$ since transformation (\r{4}) leads
to quantum mechanics, i.e. to the finite on $\s_\infty$ theory if
$u(x\in\s_\infty)=0$. It should be underlined that the theory on
$\d$-like, i.e. Dirac's, measure admits the prove {\it ex adverso}.

{\bf 6.} Eventually a necessary and sufficient condition for the
quantum field theory with symmetry in the light of GCP is knowledge
of the maximally symmetry breaking solution, $u(\bx;\q,\x,\la)$, of
Lagrange equation (\r{1}). This problem still seeks for its solution,
see also the review paper \C{21}. Perhaps it will be preferable from
the phenomenological point of view to formulate in future the theory
with symmetry directly in $W$ space, i.e. in terms of an independent
degrees of freedom. \vskip 0.5cm

{\bf Acknowledgments.} First of all I warmly remember fruitful
discussions concerning GCP with Alexei Sissakian. I am grateful to E.
Sarkisyan-Grinbaum for his useful comments to the manuscript. I am
greatly thankful to V. Kekelidze for his permanent and comprehensive
support and to V. Kadyshevsky who has encouraged me to write this
paper.

\theendnotes

\end{document}